\documentclass[twocolumn]{article}
\usepackage[paper=a4paper,margin=2cm,includehead=false]{geometry}
\usepackage[hyperfootnotes=false]{hyperref}
\usepackage{amsmath,amsfonts}
\usepackage[export]{adjustbox}
\usepackage[noend]{algpseudocode}
\usepackage{algorithmicx}
\usepackage{algorithm}
\usepackage{array}
\usepackage[caption=false,font=scriptsize,labelfont=sf,textfont=sf]{subfig}
\usepackage{textcomp}
\usepackage{stfloats}
\usepackage{url}
\usepackage{verbatim}
\usepackage{graphicx}
\usepackage[dvipsnames]{xcolor}
\usepackage{cite}
\usepackage{adjustbox}
\input{code.tex}
\usetikzlibrary{arrows.meta,intersections,shapes,arrows,calc}

\begin{document}

\title{LACIN: Linearly Arranged Complete Interconnection Networks}

\author{Ramón Beivide$^{\ast\dagger}$, Cristóbal Camarero$^\ast$, Carmen Martínez$^\ast$, Enrique Vallejo$^\ast$, Mateo Valero$^\dagger$\\
$^\ast$\textit{Universidad de Cantabria}, SPAIN\\
$^\dagger$\textit{Barcelona Supercomputing Center}, SPAIN\\
\scriptsize\{ramon.beivide, cristobal.camarero, carmen.martinez, enrique.vallejo\}@unican.es, mateo.valero@bsc.es%ORCID
}

%% mail de aceptación en 2025-12-25.
\date{This is a preprint version of a work published on Dec. 2025 at IEEE~Computer~Architecture~Letters.\\
\bgroup\scriptsize
R. Beivide, C. Camarero, C. Martinez, E. Vallejo and M. Valero, ``LACIN: Linearly Arranged Complete Interconnection Networks'' in IEEE Computer Architecture Letters, vol., no. 01, pp. 1-4, PrePrints 5555, doi: 10.1109/LCA.2025.3649284.\\
\egroup
%URL: https://doi.ieeecomputersociety.org/10.1109/LCA.2025.3649284
Can be found at \url{https://doi.org/10.1109/LCA.2025.3649284}
% https://ieeexplore.ieee.org/document/11319155
% https://www.computer.org/csdl/journal/ca/5555/01/11319155/2cSbBKtc7Ty
}

\maketitle

\begin{abstract}

Several interconnection networks are based on the complete graph topology.
Networks with a moderate size can be based on a single complete graph.
However, large-scale networks such as Dragonfly and HyperX use, respectively, a hierarchical or a multi-dimensional composition of complete graphs.

The number of links in these networks is huge and grows rapidly with their size.
This paper introduces LACIN, a set of complete graph implementations that use identically indexed ports to link switches.
This way of implementing the network reduces the complexity of its cabling and its routing.
LACIN eases the deployment of networks for parallel computers of different scales, from VLSI systems to the largest supercomputers.

\end{abstract}


\section{Introduction}

An interconnection network in which every pair of switches is connected by a link is denoted in this paper as a Complete Interconnection Network (CIN).
It has also been referred to as a fully-connected network, a full-mesh and an all-to-all network in previous works.
A CIN has the lowest possible diameter and average distance--both equal to 1--regardless the number of switches it connects, providing an advantage over any other network.

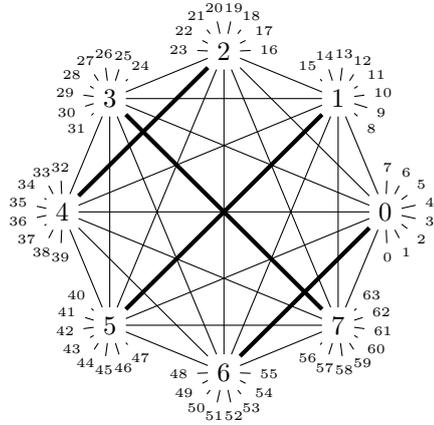
\begin{figure}
	\centering%
	\begin{tikzpicture}[scale=0.85]
	\foreach \i in {0,...,7}
	{
		\node (nodo\i) at (\i*360/8:2.5cm) {\i};
		%\foreach \j in {1,...,8} \draw (nodo\i) -- ++((\i-.5)*360/8 + .1*\j:.5cm) node {\pgfmathprintnumber{(\i-1)*8+\j}};
		\foreach \j in {0,...,7}
		{
			\path (nodo\i) ++({\i*360/8 + 25*(\j-3.5) }:.7cm) node[font=\tiny] (server) {\pgfmathparse{\i*8+\j}\pgfmathprintnumber{\pgfmathresult}};
			\draw (nodo\i) -- (server);
		}
	}
	\foreach \i in {0,...,6}
	{
		\pgfmathtruncatemacro\si{\i+1}
		\foreach \j in {\si,...,7} \draw (nodo\i) -- (nodo\j);
	}
	%%% Ahora coloreo un 1-factor. Podría parametrizarse.
	\begin{scope}[ultra thick]
	\draw (nodo7) -- (nodo3);
	\draw (nodo0) -- (nodo6);
	\draw (nodo1) -- (nodo5);
	\draw (nodo2) -- (nodo4);
	\end{scope}
	\end{tikzpicture}%
	\caption{CIN of 8 switches and 64 end-points, with 4 of its 28 links in bold.}
	\label{fig:CIN}
\end{figure}

A CIN of size $N$ is composed of $N$ switches.
Each switch has $N-1$ \textsl{network} ports to connect to every other switch.
A well-dimensioned CIN, in which every end-point can inject data at full rate under uniform random traffic, requires $N$ \textsl{edge} ports on every switch to connect computers, leading to a system with $N^2$ end-points.
Thus, the switch radix must be, at least, $2N-1$ and the total number of network links is $N(N-1)/2$.
In terms of cost, each end-point requires $1/N$ switches and $(N-1)/2N$ links.
This is, the cost in links per end-point approaches $1/2$ from below.
Figure~\ref{fig:CIN} shows a representation of a CIN for $N=8$, able to connect 64 end-point computers.
It requires a switch radix of, at least, 15 ports (8 edge ports plus 7 network ports), and $7\times 8/2=28$ wires.

The switch radix cannot be arbitrarily large, which bounds the size of a parallel computer based on a single CIN.
There are systems small enough to be interconnected using just a CIN.
Examples are NoCs~\cite{Akbari_chip}, and multi-socket systems to connect CPU and accelerator chips on server boards \cite{StarNUMA}.

For larger scales, other topologies have been proposed.
HyperX~\cite{HyperX} is a multi-dimensional network derived from the Cartesian product of complete graphs.
Its 1D version is a single CIN, while 2D and 3D variants are used in~\cite{Cascade, Intel_HyperX, kim_flat_CAL}.
The Dragonfly, employed in major supercomputers, is a two-level hierarchical network built over complete graphs~\cite{atchley2023frontier}.
Fabrics based on CINs have a high number of links, which complicates their deployment and maintenance.

This work shows that the choice of ports used to connect switches in a CIN is an important design decision.
It analyzes different alternatives, highlighting those denoted as \emph{isoport}, which only connects switches using ports with the same index.
Isoport solutions simplify wiring and routing, and favours linear implementations, denoted in this work as LACIN.
They can be applied, at least, to on-chip, multi-module chip, multi-socket, PCB, and system networks.
To the authors’ knowledge, the selection of ports for interconnecting switches has not been previously addressed.
On the one hand, when modeling networks as graphs, individual ports are not considered.
Furthermore, it is a low-level technical detail that companies do not make public disclosure.

The remainder of its content is organized as follows. Section \ref{sec:ISO} introduces the selected CIN instances. Section \ref{sec:routing} deals with their minimal routing algorithms. Section \ref{sec:LACIN} discusses LACIN alternatives. Section \ref{sec:HyperX_Df} applies LACIN to large-scale networks, and Section \ref{sec:conclusions} concludes the paper.

\section{Choice of ports and CIN instances}\label{sec:ISO}

Regardless its physical implementation, we model a CIN by means of a \emph{port pairing} matrix $P$, of $N$ rows and $N-1$ columns.
Element $P_{S, i}$ denotes the port of index $i$ in the switch $S$, with $0\leq S\leq N-1$ and $0\leq i\leq N-2$.
The $N(N-1)$ ports of the $P$ matrix are paired by $N(N-1)/2$ links according to a complete graph topology.
There is a huge number of different $P$ matrices, leading all of them to a CIN topology, but pairing their ports in different ways.
We denote each of these ways as a CIN instance.
We consider two classes: those requiring each link to pair ports with the same index, denoted as \emph{isoport}, and those allowing switches to be paired through ports with different indices, denoted as \emph{anisoport}.

\begin{figure*}[tb]
	\centering%
	\begin{tikzpicture}[scale=0.4,baseline=(label0)]
	\foreach \i in {0,...,6} \node[rotate=90,anchor=west,font=\scriptsize] at (0pt + \i*40pt,.5) {$i=\i$};
	\generalcomplete{8}{swap}\path (label0.south east) ++(120pt,0pt) node[anchor=north]{(a)};
	\end{tikzpicture}%
	\hspace{5ex plus 10ex minus 5ex}%
  	\begin{tikzpicture}[scale=0.4,baseline=(label0)]
	\def\columnhighlight{3}
	\foreach \i in {0,...,6} \node[rotate=90,anchor=west,font=\scriptsize] at (0pt + \i*40pt,.5) {$i=\i$};
	\complete{8}{circle}\path (label0.south east) ++(120pt,0pt) node[anchor=north]{(b)};
	\end{tikzpicture}%
    \hspace{5ex plus 10ex minus 5ex}%
    \begin{tikzpicture}[scale=0.4,baseline=(label0)]
	\def\columnhighlight{3}
	\foreach \i in {0,...,6} \node[rotate=90,anchor=west,font=\scriptsize] at (0pt + \i*40pt,.5) {$i=\i$};
	\complete{8}{cayley}\path (label0.south east) ++(120pt,0pt) node[anchor=north]{(c)};
	\end{tikzpicture}%
    \caption{Port Pairing $P$ Matrices with $N=8$ for (a) anisoport Swap, (b) isoport Circle and (c) isoport XOR CIN instances. $1$-factor $i=3$ in bold.}\label{fig:K_8}
\end{figure*}
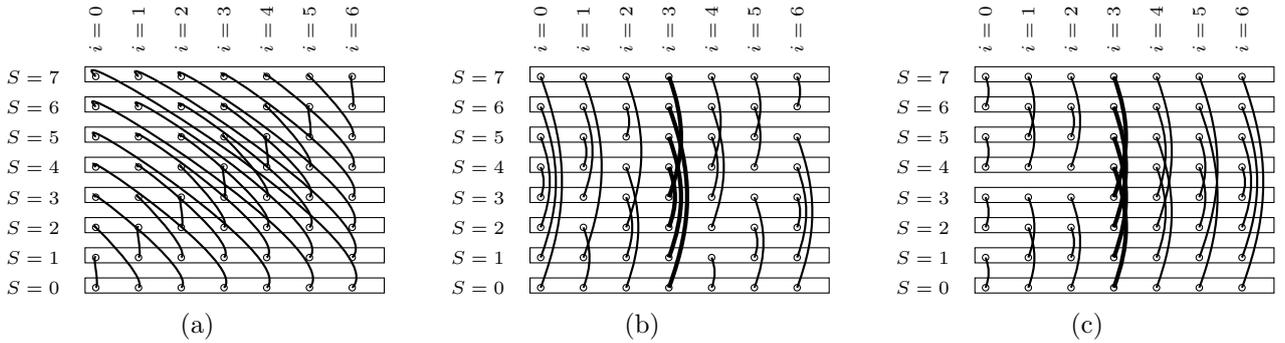

An anisoport CIN instance can be naturally obtained by successively connecting each switch to all the others, using the first available ports on each of the switches.
With this approach, denoted as \textsl{Swap} in this work, port $P_{S, i}$ is paired to port $P_{i+1, S}$ when $S\leq i$, or to port $P_{i, S-1}$ when $S>i$.
Figure \ref{fig:K_8} (a) shows the Swap $P$ matrix for a CIN of 8 switches.

No known isoport CIN instances have been studied in the literature.
This work introduces two alternatives based on complete graph factorizations.
A $1$-factor of an even order graph is a \textsl{perfect match}, that is, a pairing of its $N$ vertices by means of $N/2$ edges.
For $N$ even, a $1$-factorization has $N-1$ $1$-factors, the same as the number of network ports per switch.
Isoport instances use the $N$ ports of index $i$ to build the $1$-factor $i$, with $0\leq i\leq N-2$.
The next two factorizations, long used in step-wise all-to-all algorithms to reduce congestion \cite{Bokhari_exchange}, \cite{Schmiermund}, are customized here to define isoport CIN instances.

The first proposed one, denoted as \emph{Circle}, adapts an algorithm used in round-robin tournaments to pair the contenders.
This CIN instance comes from Algorithm~\ref{algo:circle} and its $P$ matrix for $N=8$ can be seen in Figure \ref{fig:K_8} (b).
Although odd-sized graphs cannot be $1$-factorized, a Circle $P$ matrix can be obtained from that of the even case ($N+1$) by removing the last row and its incident links.
One port per switch remains idle, but the isoport property is preserved.

The second proposal, denoted as \emph{XOR}, can be used when $N=2^{n}$.
It is based on the bit-wise exclusive OR function, $\oplus$, applied to the switch labels of $n$ bits.
If $A$ and $B$ are the labels of two switches, the index of the two ports that link them is $i = A \oplus B - 1$.
Since the XOR is self-inverse, if $i = A \oplus B -1$, then $B = A \oplus (i +1)$, meaning that switch $B$ is the neighbor of switch $A$ reachable through port of index $i$.
Thus, in general, the port $P_{S, i}$ is connected to port $P_{S\oplus (i + 1), i}$.
Figure \ref{fig:K_8} (c) shows the XOR $P$ matrix for a CIN with 8 switches.

\newcommand\var[1]{\ensuremath{\mathtt{\detokenize{#1}}}}
\begin{algorithm}[t!]
	\small
	\begin{algorithmic}
		\Require{Even \var{N}}
		%\Ensure{A $\var{N} \times (\var{N}-1)$ \var{Matrix} with port offset values}
		\For{\var{i} from 0 to $\var{N}-2$}
			\For{\var{S} from 0 to $\var{N}-1$}
				\If{$\var{S}=\var{N}-1$}
					\State \textcolor{ForestGreen}{\% Connect the last node to the rest of nodes.}
					\State Connect port $P_{S, i}$ to port $P_{i, i}$.
				\ElsIf{$\var{S}=\var{i}$}
					\State \textcolor{ForestGreen}{\% Any node connected to the last node.}
					\State Connect port $P_{S, i}$ to port $P_{N-1, i}$.
				\Else
					\State \textcolor{ForestGreen}{\% Parallel connections.}
					\State Connect port $P_{S, i}$ to port $P_{(2i-S)\bmod (N-1), i}$.
				\EndIf
			\EndFor
		\EndFor
	\end{algorithmic}
	\caption{\emph{Circle} LACIN for $N$ even. Determines the neighbours of every switch through its ports $P_{S, i}$.}\label{algo:circle}
\end{algorithm}

Regardless of the physical arrangement of the switches, isoport CIN instances offer several advantages.
They minimize mistakes during network deployment and maintenance.
First, links and ports are classified using $N-1$ colours, facilitating cable organization.
Second, a link ending on a certain port in a switch, must also end in a port with the same index in other switch, which reduces uncertainty.
Moreover, as next section shows, they enable simple routing algorithms.

\section{Minimal Routing}\label{sec:routing}

In a CIN, a computer $C$ has a global address of two digits, that is, $C = (C_1,C_0)$, with $C_1$ identifying the switch and $C_0$, the end-point computer in that switch.
In Figure \ref{fig:CIN}, $C_1$ corresponds to the three high-order bits of the binary label of $C$, and $C_0$, to the three low-order bits. Suppose a source computer $A = (A_1,A_0)$ sending data to a destination computer $B = (B_1,B_0)$ through a minimal path.
Regardless the routing employed, if the two end-points are connected to the same switch, $A_1 = B_1$, or the packet has arrived to the destination switch after a network hop, the port indexed by $B_0$ is used to eject the packet.

%The port pairing matrix of a CIN instance induces its minimal routing algorithm.
When $A_1 \neq B_1$, the port that must be used at the switch $A_1$ to send data to $B_1$ is determined by the CIN instance.
The port index used by Swap is $i=B_1$ if $A_1 \leq B_1$ or $i=B_1+1$ otherwise. In XOR, it is the one indexed by $i= A_1 \oplus B_1 -1$.
The port selected by Circle is described by Algorithm~\ref{algo:routing_circle}, which has been derived from Algorithm~\ref{algo:circle}.
It requires to consider as special cases the routes from, or to, the $N-1$ switch, for which it has several branches with different comparisons/additions.

The simplest routing corresponds to XOR, which relies solely on logic gates plus a decrementer.
In Swap, routing is a little bit more costly as it requires a comparator besides an incrementer.
In Circle, it is more complex than in the others, but it is still simple enough to be implemented directly in hardware, avoiding the use of routing tables.
These routings can be extended to higher dimensional networks.

In a CIN with minimal routing all the paths are of length one and deadlock cannot occur.
The previous algorithms can be employed as the basis of adaptive mechanisms that use either minimal or non-minimal paths depending on the traffic nature.
In such a case, deadlock must be avoided either restricting non-minimal routes or by using two virtual channels.

\begin{algorithm}[t!]
	\small
	\begin{algorithmic}
	%	\Require{Even number of switches \var{N}.}
		\Require{Even number of switches \var{N}, Source switch \var{A}, Destination switch \var{B}, $0\leq \var{A},\var{B}<N-1$, $\var{A}\neq\var{B}$.}
		\Ensure{\var{i}, $0\leq \var{i}\leq N-2$, such that port \var{i} of \var{A} connects to \var{B}.}
		\State $\var{T}\leftarrow \var{A}+\var{B}$
		\If{$\var{T}=\var{N}-1$}
			\State $\var{i}\leftarrow 0$
		\ElsIf{$\var{B}=\var{N}-1$}
			\State $\var{i}\leftarrow \var{A}$
		\ElsIf{$\var{A}=\var{N}-1$}
			\State $\var{i}\leftarrow \var{B}$
		\ElsIf{\var{T} is even}
			\State $\var{i}\leftarrow \var{T}/2$
		\ElsIf{$\var{T}<\var{N}-1$}
			\State $\var{i}\leftarrow (\var{T}+\var{N}-1)/2$
		\Else
			\State $\var{i}\leftarrow (\var{T}-\var{N}+1)/2$
		\EndIf
	\end{algorithmic}
	\caption {Packet Routing for \emph{Circle} with $N$ even.}\label{algo:routing_circle}
\end{algorithm}

\section{LACIN: Linear Arranged CIN}\label{sec:LACIN}

LACIN is an isoport CIN instance implementation whose switches are arranged linearly.
It is motivated by designs where switches are placed in a straight line with links running parallel, minimizing crossings.
This layout can address various requirements, including implementation medium, large-scale network use, ease of deployment, cooling, repair, and maintenance.
Examples include the on-chip CIN in~\cite{Akbari_chip} and Intel PIUMA~\cite{Intel_HyperX}, which employs a HyperX topology.

Coming LACIN from an isoport CIN instance, all its ports are connected within columns.
This layout results in straight links between switches, which reduces their total length compared to any other 1D implementation.
LACIN requires wires of different lengths, up to the number of network ports per switch. It needs $w$ wires of length $N-w$, with $1\leq w \leq N-1$, which results a total wire length of $(N^{3}-N)/6$.

On the contrary, linear anisoport implementations require oblique cables, which means greater length.
For simplicity, we assume that distances between switches and ports are similar.
In particular, with Swap, links with a vertical length of $k$ have a horizontal length of $k-1$.
Thus, the total length is approximately $\sqrt{2}=1.41$ times the total length of a LACIN.

Depending on the deployment medium, such as in VLSI or on-board systems, avoiding wire crossings is pursued.
In LACIN, wires connecting ports with different indices belong to different $1$-factors, and cannot cross.
As shown next, in Circle, the $N/2$ links within each $1$-factor do not cross either.

The $1$-factor $i$, with $0\leq i\leq N-2$, is induced by the connection between node $i$ and node $N-1$.
Any $1$-factor is composed of, at least, $(N/2)-1$ parallel links that can be crossed by the single link that connects nodes $i$ and $N-1$.
This single link can cross either all or some or none of the parallel links.
In Figures \ref{fig:CIN} and \ref{fig:K_8} (b), the links belonging to the $1$-factor $i=3$ are highlighted.
It has three parallel links crossed by the link that connects switches 3 and 7.
In $1$-factors $i=0$ and $i=6$ all the links are parallel as nodes $0$ and $6$ are neighbours of node $7$ in Figure \ref{fig:CIN}.
In general, for the $1$-factor $i$, the possible crossing link traverses $i$ parallel links when $0\leq i\leq (N/2)-1$, and $N-2-i$ parallel links when $N/2\leq i\leq N-2$.
If all the parallel wires of any $1$-factor $i$ run to the right of the port $i$ and the wire that can cross them runs to its left, the resulting layout has not crossings at all.

Similar layouts for XOR LACIN are more complex.
The simple drawings used in this paper result in wire crossings, which number increases with the network size.
Observe, for example, the $1$-factor $i=3$ highlighted in Figure \ref{fig:K_8} (c).
Its crossovers can be avoided by introducing turns in some wires, but the resulting layout would not be as simple as Circle’s.

\begin{table}
	\centering%
	%\caption{Summary of properties of different CIN layouts: the number of switches, the asymptotic total wire length (normalized to optimal) and the routing cost in terms of the number of addition and comparison of two registers operations.}\label{tbl:resumen}
	\caption{Summary of properties of different 1D CIN layouts.}\label{tbl:resumen}
	\scriptsize
	%\begin{tabular}{|l|llp{15ex}|}
	\renewcommand{\thempfootnote}{\fnsymbol{mpfootnote}}%Que es mejor para esto, letras (default \alph) o símbolos?
	\begin{minipage}{\linewidth}
	\let\footnoterule\relax%Remove the rule separating the footnotes.
	\setlength\tabcolsep{4pt}%default 6pt
	\begin{tabular}{|l|l|l|l|l|}
	\hline
	CIN Layout & Isoport & $\#$ Switches	&  Wire length\footnote{Total asymptotically wire length normalized to minimum.}	& Routing cost\footnote{Additional number of adders and/or comparators with respect to XOR.}\\
	\hline
	Swap	&	No& Any		&	$\sqrt 2$	&	1\\
	Circle	&	Yes & Any		&	1	&	5\\
	XOR		&	Yes & $N=2^n$	&	1	&	---\\
	\hline
	\end{tabular}
    \vspace{-1em}
	\end{minipage}
\end{table}

The main properties of the linear deployments of the different CIN instances are summarized in Table \ref{tbl:resumen}.

\begin{figure}[tbh]
	\centering%
	\begin{tikzpicture}[scale=0.8]
	%% Draw major links. Part of them will be overdrawn later.
	\foreach \majori in {0,1,2}
	{
		\pgfmathtruncatemacro\xmajori{\majori/2}
		\pgfmathtruncatemacro\ymajori{mod(\majori,2)}
		\pgfmathtruncatemacro\start{\majori+1}
		\foreach \majorj in {\start,...,3}
		{
			\pgfmathtruncatemacro\xmajorj{\majorj/2}
			\pgfmathtruncatemacro\ymajorj{mod(\majorj,2)}
			\draw[ultra thick,name path global={major_\majori_\majorj}] (\xmajori*3+.5,\ymajori*3+.5) -- (\xmajorj*3+.5,\ymajorj*3+.5);
		}
	}
	\foreach \major in {0,1,2,3}
	{
		\pgfmathtruncatemacro\xmajor{\major/2}
		\pgfmathtruncatemacro\ymajor{mod(\major,2)}
		\draw[thick,fill=white,rounded corners,name path global=major_\major] (\xmajor*3,\ymajor*3) ++(-.5,-.5) rectangle ++(2,2);%Draw white rectangles above the major links.
		\foreach \minor in {0,1,2,3}
		{
			\pgfmathtruncatemacro\xminor{\minor/2}
			\pgfmathtruncatemacro\yminor{mod(\minor,2)}
			\pgfmathtruncatemacro\index{\major*4+\minor}
			%% The nodes with their numbers.
			\node[draw,minimum width=15pt,minimum height=15pt,inner sep=0pt] (nodo_\major_\minor) at (\xmajor*3+\xminor,\ymajor*3+\yminor) {\pgfmathprintnumber{\index}};
		}
		%% Draw the minor links.
		\foreach \minori in {0,1,2}
		{
			\pgfmathtruncatemacro\start{\minori+1}
			\foreach \minorj in {\start,...,3}
				\draw (nodo_\major_\minori) -- (nodo_\major_\minorj);
		}
	}
	%% lines into the bundles.
	\begin{scope}[densely dashed,ultra thin]
	\foreach \majori in {0,1,2}
	{
		\pgfmathtruncatemacro\start{\majori+1}
		\foreach \majorj in {\start,...,3}
		{
			\path [name intersections={of={major_\majori} and {major_\majori_\majorj}, by=joint}];
			\foreach \minor in {0,1,2,3}
			{
				\pgfmathtruncatemacro\cond{
					\majori+\majorj==3 && (\minor==\majori || \minor+\majori==3)
					%% It is a diagonal major and the minor is at the same corner as the major.
					%% We also change the minor in the opposite side.
				}
				\ifthenelse{\cond=1}
				{
					\pgfmathtruncatemacro\cond{\minor==\majori}
					\ifthenelse{\cond=1}
					{\pgfmathsetmacro\angle{90*(\minor==0?0:mod(\minor+1,3)+1)+30}}
					{\pgfmathsetmacro\angle{90*(\minor==0?0:mod(\minor+1,3)+1)+240}}
					\draw[] (nodo_\majori_\minor.\angle) -- (joint);
				}{
					\draw[] (nodo_\majori_\minor) -- (joint);
				}
			}
			\path [name intersections={of={major_\majorj} and {major_\majori_\majorj}, by=joint}];
			\foreach \minor in {0,1,2,3}
			{
				\pgfmathtruncatemacro\cond{
					\majori+\majorj==3 && (\minor==\majorj || \minor+\majorj==3)
				}
				\ifthenelse{\cond=1}
				{
					\pgfmathtruncatemacro\cond{\minor==\majorj}
					\ifthenelse{\cond=1}
					{\pgfmathsetmacro\angle{90*(\minor==0?0:mod(\minor+1,3)+1)+30}}
					{\pgfmathsetmacro\angle{90*(\minor==0?0:mod(\minor+1,3)+1)+240}}
					\draw[] (nodo_\majorj_\minor.\angle) -- (joint);
				}{
					\draw[] (nodo_\majorj_\minor) -- (joint);
				}
			}
		}
	}
	\end{scope}
	\end{tikzpicture}
	\caption{A 2D CIN implementation of size 16 as a hierarchy of CINs of size 4. The links of the outer CIN are bundles, each comprising 16 inner links.}
	\label{fig:CIN-on-CIN}
    \vspace{-1em}
\end{figure}
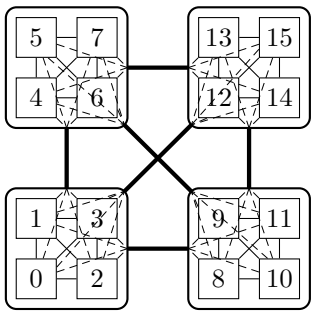

Alternatively, the switches of a CIN, being it isoport or not, can be arranged in two physical dimensions.
For example, for $N=16$, a 2D deployment based on 4 partitions of 4 switches each can be seen in Figure~\ref{fig:CIN-on-CIN}.
Its 120 links are organized in two levels: $24 = 4 \times 6$ links inside the partitions and 96 links between partitions, grouped in 6 hoses of 16 wires each.

Identical scheme, but with 8 partitions, is used to implement groups in HPE Dragonflies \cite{atchley2023frontier}.
They are distributed inside the racks as 2 columns of 4 partitions, needing 28 bundles of 16 wires to fully connect the 8 partitions.
This 2D layout is more complex than LACIN as it needs diagonal links at two levels and entails wire crossings.
Maintaining such scheme could incur in higher costs and potential issues to isolate and replace malfunctioning links.
In addition, the total length of wire, when considering deployments including rack space for the end-point computers, is similar in both cases.

As in both cases $N=2^{n}$, the previous 2D implementations could perfectly be based on XOR CIN instances and beneficiate from its wire organization and routing.

\section{LACIN for Large-scale Networks}\label{sec:HyperX_Df}

LACIN can be applied to multi-dimensional HyperX networks.
Let us consider as an example a XOR LACIN implementation of a diameter $3$ network, with 65,536 end-points and 4,096 switches arranged as $16 \times 16 \times 16$ HyperX.
Switches employ 61 ports: 16 edge ports for end-points and 15 network ports per dimension to connect other switches.
%The system has 92,160 network links.
An end-point computer, $C$, has a 16-bit address composed of 4 digits of 4 bits each.
The three high-order digits, $C_3,C_2,C_1$, identify the coordinates $Z, Y, X$ of the switch, and the low-order digit $C_0$, the specific computer connected to this switch.

\begin{figure}[tb]
	\centering%
	\begin{tikzpicture}[scale=0.75, x=1.7cm,y=1.7cm]
	%\begin{scope}[yshift=10cm]
	%	\hamming{4}{cayley}{4}{cayley}
	%	\node[draw,double=red!50!white,thick,fill=white] at (2,2) {Rehacer sin cruces};
	%\end{scope}
	%% __ Y ahora - sin cruces. __ %%
	\pgfmathsetmacro{\routersize}{.8}%
	\pgfmathsetmacro{\portlineoffset}{.08}%
	\pgfmathsetmacro{\firstportoffset}{.2}%
	\pgfmathsetmacro{\horizontalportsize}{(\routersize-\firstportoffset)/(3)}%
	\pgfmathsetmacro{\verticalportsize}{(\routersize-\firstportoffset)/(3)}%
	\tikzset{port/.style={circle,draw,inner sep=0,minimum width=3pt}}
	\foreach \horizontalindex in {0,...,3}
	{
		\foreach \verticalindex in {0,...,3}
		{
			%\draw (\horizontalindex,\verticalindex) ++(-.4,-.4) rectangle ++(.8,.8);
			\draw (\horizontalindex,\verticalindex) rectangle ++(\routersize,\routersize);
		}
	}
	\foreach \horizontalindex in {0,...,3}
	{
		\foreach \verticalindex in {0,...,3}
		{
			%horizontal lines
			\foreach \portindex in {0,...,2}
			{
				\path  (\horizontalindex,\verticalindex) ++(\portlineoffset,\firstportoffset+\verticalportsize*\portindex) node[port] (puerto H \horizontalindex\verticalindex\portindex) {};
				%\pgfmathtruncatemacro{\destination}{\hamming@horizontal@function(\horizontalindex,\portindex,\hamming@horizontal@size)}%
				%\pgfmathtruncatemacro{\cond}{\destination>\horizontalindex && \destination<\hamming@horizontal@size}
				%\ifthenelse{\cond=1}{
				%	\lanestyle
				%	%\draw[lane,thick] (\horizontalindex,\verticalindex) ++(\portlineoffset,0) -- ++(2pt+1pt*\portindex,1pt + 2.0pt*\lacin@lane + 1pt*\portindex) -| (\lacin@target,\verticalindex);
				%	\path (\horizontalindex,\verticalindex) ++(\portlineoffset,\firstportoffset+\verticalportsize*\portindex) coordinate (s) ++(0,2pt + 2.0pt*\lacin@lane ) coordinate(sm);
				%	\path (\lacin@target,\verticalindex) ++(\portlineoffset,\firstportoffset+\verticalportsize*\portindex) coordinate (e) ++(0,2pt + 2.0pt*\lacin@lane ) coordinate(em);
				%	%\draw[lane,thick] (s) -- (sm) -- (em) -- (e);
				%	\draw[lane,thick] (s) edge[out=10,in=170] (e);
				%}{}
			}
			%vertical lines
			\foreach \portindex in {0,...,2}
			{
				\path  (\horizontalindex,\verticalindex) ++(\firstportoffset+\horizontalportsize*\portindex,\portlineoffset) node[port] (puerto V \horizontalindex\verticalindex\portindex) {};
				%\pgfmathtruncatemacro{\destination}{\hamming@vertical@function(\verticalindex,\portindex,\hamming@vertical@size)}%
				%\pgfmathtruncatemacro{\cond}{\destination>\verticalindex && \destination<\hamming@vertical@size}
				%\ifthenelse{\cond=1}{
				%	\lanestyle
				%	%\draw[lane,thick] (\horizontalindex,\verticalindex) ++(0,\portlineoffset) -- ++(1pt + 2.0pt*\lacin@lane + 1pt*\portindex,2pt+1pt*\portindex) |- (\horizontalindex,\lacin@target);
				%	\path (\horizontalindex,\verticalindex) ++(\firstportoffset+\horizontalportsize*\portindex,\portlineoffset) coordinate (s) ++(2pt + 2.0pt*\lacin@lane ,0) coordinate (sm);
				%	\path (\horizontalindex,\lacin@target) ++(\firstportoffset+\horizontalportsize*\portindex,\portlineoffset) coordinate (e) ++(2pt + 2.0pt*\lacin@lane ,0) coordinate (em);
				%	%\draw[lane,thick] (s) -- (sm) -- (em) -- (e);
				%	\draw[lane,thick] (s) edge[out=80,in=-80] (e);
				%}{}
			}
		}
	}

	\foreach \verticalindex in {0,...,3}
	{
		\draw[thick] (puerto H 0\verticalindex 0) edge[out=10,in=170]   (puerto H 1\verticalindex 0);
		\draw[thick] (puerto H 0\verticalindex 1) edge[out=10,in=170]   (puerto H 2\verticalindex 1);
		\draw[thick] (puerto H 0\verticalindex 2) edge[out=10,in=170]   (puerto H 3\verticalindex 2);
		\draw[thick] (puerto H 1\verticalindex 1) edge[out=-10,in=-170] (puerto H 3\verticalindex 1);
		\draw[thick] (puerto H 1\verticalindex 2) edge[out=10,in=170]   (puerto H 2\verticalindex 2);
		\draw[thick] (puerto H 2\verticalindex 0) edge[out=10,in=170]   (puerto H 3\verticalindex 0);
	}
	\foreach \horizontalindex in {0,...,3}
	{
		\draw[thick] (puerto V \horizontalindex 00) edge[out=80,in=-80]   (puerto V \horizontalindex 10);
		\draw[thick] (puerto V \horizontalindex 01) edge[out=80,in=-80]   (puerto V \horizontalindex 21);
		\draw[thick] (puerto V \horizontalindex 02) edge[out=80,in=-80]   (puerto V \horizontalindex 32);
		\draw[thick] (puerto V \horizontalindex 11) edge[out=100,in=-100] (puerto V \horizontalindex 31);
		\draw[thick] (puerto V \horizontalindex 12) edge[out=80,in=-80]   (puerto V \horizontalindex 22);
		\draw[thick] (puerto V \horizontalindex 20) edge[out=80,in=-80]   (puerto V \horizontalindex 30);
	}

	\end{tikzpicture}
	\caption{Top view of a 4$\times$4$\times$4 HyperX using XOR LACINs. Each rack stacks 4 switches. Only inter-rack connections, grouped into super-ports, are shown.}\label{fig:HX}
    \vspace{-1em}
\end{figure}
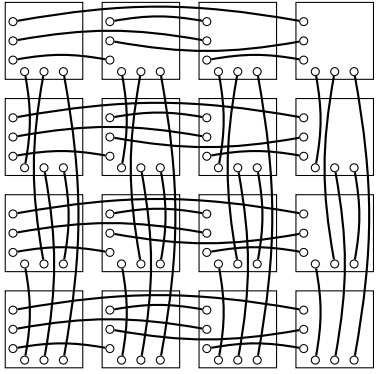

Assume that a chassis contains one switch and their corresponding 16 end-point computers.
Provided that a rack stacks 16 chassis, a network dimension, let us say $Z$, is implemented inside the racks using LACIN.
The 120 $Z$  links run along the vertical dimension of the rack, and are organized in 15 columns of 8 wires each.
The system is a square of $16 \times 16$ racks connected by LACINs to implement the two other network dimensions in the $XY$ plane.
Racks have 15 super-ports for dimension $X$ and other 15 for dimension $Y$, each grouping 16 individual ports.
There are 120 hoses of 16 wires each for linking the super-ports of each row or column.
Super-ports and hoses are labelled with 15 colours, which allows to organize hoses in 15 classes of 8 hoses each.
Figure \ref{fig:HX} shows such deployment for a small $4 \times 4\times 4$ HyperX.

To illustrate routing, assume a source computer $A = (A_3,A_2,A_1,A_0)$ sending a packet to a target computer $B = (B_3,B_2,B_1,B_0)$ through a minimal path.
It will employ port $P_{A_3 \oplus B_3-1}$ at the source switch to route trough dimension $Z$.
Ports $P_{A_2 \oplus B_2-1}$ and $P_{A_1 \oplus B_1-1}$ are used for the $Y$ and $X$ dimensions, respectively, and port $P_{B_0}$ at the target switch to deliver the packet.
When the XOR of the source and destination digits of a given dimension returns 0, that dimension is skipped.
The order in which the packet crosses dimensions can be either deterministic, such as DOR, or adaptive.
While DOR in HyperX does not require virtual channels (i.e., only one buffer per port, or just 1 physical channel) adaptive mechanisms typically require them to avoid deadlock.
Adaptive non-minimal algorithms can be easily derived from this minimal scheme.

Dragonflies connect switch groups via a global CIN, with each group wired as a local CIN.
Currently, group size is limited by electrical wire range, resulting in one-rack groups that can be implemented using a LACIN along the rack’s vertical dimension.
Applying LACIN to the global network induces linear rack organizations.
Co-packaged silicon photonics removes this limitation, enabling larger groups arranged in rack rows, with LACIN used horizontally for local CINs.
Each rack switch dedicates some ports to the global LACIN, which connects, column-wise, rows of global ports.

\section{Conclusions}\label{sec:conclusions}

The relevance of the complete graph among current interconnection networks deserves its detailed analysis.
This work presents different alternatives to connect the ports of a CIN, which are independent of their implementations but condition wiring organization and routing algorithms.
It proposes a way to connect switches in which every link employs the same port index at both endpoints.
Based on this, it introduces LACIN, a specific linear CIN implementation.
LACIN offers advantages, including its ability to organize links and ports, which facilitates the deployment of systems based on a single CIN and that of large networks, such as HyperX and Dragonfly.
In addition, LACIN minimizes the length of wiring required by any linear CIN implementation.

LACIN implementations show interesting trade-offs.
While Circle can be defined for any network size, XOR only can be used in networks whose number of switches per dimension is a power of 2.
Circle leads to the simplest implementation for avoiding wire crossing.
With respect to routing, although both can avoid the use of tables, it is simpler in XOR.

\section*{Acknowledgments}
This work has been supported by the Spanish Ministry
of Science and Innovation under contracts PID2019-
105660RB-C22, TED2021-131176B-I00, and PID2022-
136454NB-C21. C. Camarero is supported by the Spanish
Ministry of Science and Innovation, Ramón y Cajal
contract RYC2021-033959-I. R. Beivide is supported
by The Barcelona Supercomputing Center (BSC) under
contract CONSER02023011NG.

%R. Beivide is supported by The Barcelona Supercomputing Center (BSC) under contract CONSER02023011NG.
%C. Camarero is under Ramón y Cajal contract RYC2021-033959-I from Spain's Ministerio de Ciencia e Innovación with funding from the Mecanismo de Recuperación y Resiliencia de la Unión Europea.
%The authors would like to thank these projects: PLANIFICADORES Y REDES PARA DATA CENTERS SOSTENIBLES project TED2021-131176B-I00 with funding from MCIN/ AEI /10.13039/501100011033 and Unión Europea/NextGenerationEU/PRTR; REDES DE INTERCONEXIÓN, ACELERADORES HARDWARE Y OPTIMIZACIÓN DE APLICACIONES, project PID2019-105660RB-C22 with funding from MCIN/ AEI /10.13039/501100011033; and ARQUITECTURA Y PROGRAMACIÓN DE COMPUTADORES ESCALABLES DE ALTO RENDIMIENTO Y BAJO CONSUMO III-UC (TEAM-MATES UC) project PID2022-136454NB-C21 with funding from MICIU/AEI /10.13039/501100011033 and FEDER, UE.

\bibliographystyle{plain}
\bibliography{main}

\vfill

\end{document}